\documentclass[preprint,showpacs,preprintnumbers,amsmath,amssymb]{revtex4}
\usepackage{dcolumn}
\usepackage{graphicx}
\usepackage{bm}

\begin{document}
\title{Helical Disruptions in Small Loops of DNA}

\author{Marco Zoli}

\affiliation{School of Science and Technology - CNISM \\  University of Camerino, I-62032 Camerino, Italy \\ marco.zoli@unicam.it}

\begin{abstract}
The thermodynamical stability of DNA minicircles is investigated by means of path integral techniques. Hydrogen bonds between base pairs on complementary strands can  be broken by thermal fluctuations and temporary fluctuational openings along the double helix are essential to biological functions such as transcription and replication of the genetic information.  Helix unwinding and bubble formation patterns are computed in circular sequences with variable radius in order to analyze the interplay between molecule size and appearance of helical disruptions. The latter are found in minicircles with $< 100$ base pairs and appear as a strategy to soften the stress due to the bending and torsion of the helix.
\end{abstract}

\maketitle

It is known that the helicoidal conformation of DNA is essentially determined by the hydrophobicity of purine and pyrimidine bases and by the bond angles in the flexible sugar-phosphate backbone while sequence of the bases and environmental conditions due to the solvent also contribute to the molecule shape \cite{calla}. As each strand bears a negative charge ($e$) for each phosphate group and the rise distance between adjacent nucleotides is $\sim 3.4$\AA, the bare double helix has a high linear charge density of $\sim 0.6 \,e^{}\, \AA^{-1}$. Although the effective charge density is reduced by the counterions in the solvent, the electrostatic strands repulsion is key to the stability of the helix and also affects the inter-helical chiral interactions in those condensed phases of DNA assemblies (such as liquid crystals) which underlie the impressive growth of DNA-based structures recently witnessed in materials science.

Base pairing and base stacking are the fundamental interactions which control the synthesis of DNA and determine the thermodynamic stability both of single helices and of helix aggregates. However even stable duplexes at room temperature show local openings, temporary bubbles, which are intrinsic to the biological functioning as they permit the transcription and replication of the genetic code. Such bubbles are due to the strong fluctuational effects on the hydrogen bonds between complementary strands and cause the local unwinding of the double helix which ultimately leads to a state of negative supercoiled DNA for almost all living beings.
While these processes are qualitatively understood, quantitative predictions of (energetically) optimal helical configurations for specific systems are scarce. We contribute to fill this gap by introducing a new path integral computational method which readily applies to loops of DNA as those found in bacterial plasmids, viral genomes and also mammalian cells.

\section{Mesoscopic Model }

\begin{figure}
\includegraphics[height=7.5cm,width=12.5cm,angle=-90]{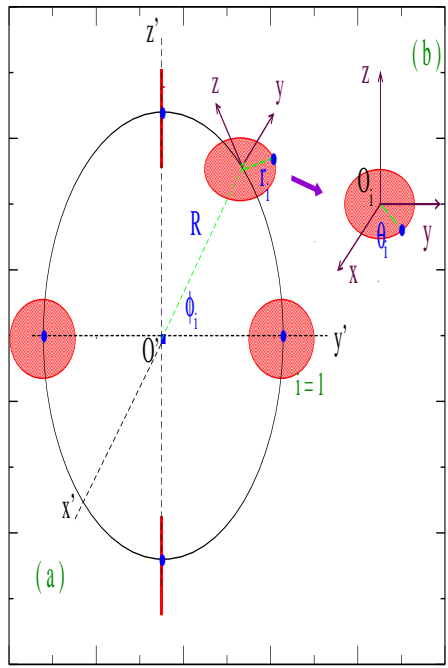}
\caption{\label{fig:1}(Color online) (a) Helicoidal model for circular DNA with bending planes. The blue filled circles are the centers of mass of the base pairs stacked along the molecule backbone with a rise distance $3.4$ \AA. In the ground state all \emph{bps} lie on the circumference with radius $R$. The red-shaded areas are spanned by the fluctuational vectors whose amplitude is measured by $|r_i|$ for the $i-$ \emph{bp}.  The azimuthal angle $\phi_i$ measures the bending of the $i-$ \emph{bp} plane with respect to the $(x',y')$ plane, $\textbf{x'}$ being normal to the sheet plane. (b) Local reference system for the $i-$ \emph{bp}. $\theta_i$ is the twist around the molecule backbone. The z-axis is tangent to the ground state circle.}
\end{figure}

Let's arrange $N$ base pairs (\textit{bps}) on a circle with radius $R$ such that, $2 \pi R / N \sim 3.4$ \AA, as depicted in Fig.~\ref{fig:1}.
When all \textit{bps} centers of mass lie on the circumference, which represents the molecule backbone, the system is in the ground state.

Say $\textbf{r}_i$, the inter-strand fluctuation for the \textit{i}-base pair ($i=1,..N$) with respect to the ground state.  
Hence, we define the  vector $\textbf{t}_i$:

\begin{eqnarray}
\bigl({t }_i \bigr)_{x} =\, |\textbf{r}_i| \cos\phi_i \cos\theta_i ;  \, \, \, \, \bigl({t }_i\bigr)_{y} =\,(R + |\textbf{r}_i|\sin\theta_i) \cos\phi_i
;  \, \, \, \, \bigl({t }_i\bigr)_{z} =\,(R + |\textbf{r}_i|) \sin\phi_i \,.
\label{eq:004}
\end{eqnarray}

The ground state is recovered once all \textit{bps}-fluctuations vanish hence, $t_i =\,R ,\,\, \forall i$.
The polar angle, $\theta_i =\, (i - 1) 2\pi / h + \theta_S$, measures the $i-$ \emph{bp} twisting around the molecule backbone, with $h=\, N / Tw$ being the number 
of \textit{bps} per helix turn and $Tw$ is the  twist number accounting for the coiling of the individual strands around the helical axis \cite{calla}. 
The azimuthal angle, $\phi_i =\, (i-1){{2 \pi} / N} + \phi_S $,  defines the bending between adjacent \emph{bps} along the stack.
As the polynucleotide chain has a direction due to the chemistry of the intra-strand bonds, a distribution of values for
the twist ($\theta_S$) and the bending ($\phi_S $) of the first \emph{bp} in the sequence is weighed in the computation.
The fluctuational orbits defined by $i=\,1$ and $i=\,N+1$ overlap consistently with the closure condition holding for the DNA ring.

\section{Space-Time Mapping}

My previous path integral analysis of DNA \cite{io09,io10,io11a,io11b,io12,io13a,io13b,io14a,io14b} were based on the ansatz that
the \emph{bps} displacements could be treated as one dimensional paths $x(\tau_i)$,  $|\textbf{r}_i| \rightarrow \, x(\tau_i)$,  with the imaginary time $\tau_i \in [0, \beta]$ and $\beta$ being the inverse the temperature. 

Here I introduce a more general (albeit more CPU time consuming) space-time mapping technique,  which does not pin a base pair to a specific $\tau_i$ 
thus avoiding the somewhat arbitrary partition of the $\beta$ length in $N$ intervals:

\begin{eqnarray}
|\textbf{r}_i| \rightarrow  x_i(\tau) ;  \, \, \, \,  |\textbf{t}_i| \rightarrow  \eta_i(\tau)  ;  \, \, \, \, \,  \tau \in [0 \,, \beta ] \,.
\label{eq:005}
\end{eqnarray}

The paths $x_i(\tau)$ are expanded in Fourier series, the constraint $x_{i=\,1}(\tau)=\,x_{i=\,N+1}(\tau)$ for a DNA loop is implemented and, for the model in Fig.~\ref{fig:1}, the\textit{ bps} fluctuations in the path integral formalism are given by:

\begin{eqnarray}
& & \eta_i(\tau)=\, \bigl[R^2 + x_i(\tau)^2 + 2R |x_i(\tau)|f(\theta_i,\phi_i) \bigr]^{1/2} \, \nonumber
\\
& &f(\theta_i,\phi_i)=\,\sin\theta_i \cos^2 \phi_i + \sin^2 \phi_i \, . \,
\label{eq:006}
\end{eqnarray}

This new mapping technique lets $\tau$ as an integration variable and permits to set a realistic rise distance between adjacent nucleotides along the stack. 
Written in terms of the $\eta_i(\tau)$, the Hamiltonian contains: \textit{ i)} a Morse potential, $V_M[\eta_i(\tau)]$, describing the effective hydrogen bond interaction between \textit{bps} mates, \textit{ ii)} a solvent term, $V_{sol}[\eta_i(\tau)]$, accounting for hydrogen bond recombination with the counterions dissolved in water, \textit{ iii)} a (two particles) stacking potential, $V_{S}\bigl[ \eta _i(\tau), \eta _{i-1}(\tau) \bigr]$, between adjacent bases along the strand. The potential parameters are widely discussed in Refs. \cite{io11a,io11b,io12}.
Accordingly, the classical action $A[\eta_{i} ]$ is a $\tau$-integral of the base pair \textit{kinetic \textit{plus} potential} energies \cite{io14a} and the partition function $Z_N$ for a sequence with $N$ \textit{bps} reads:

\begin{eqnarray}
& &Z_N =\,\prod_{i=1}^{N} \oint {D}x_{i} \sum_{\theta_S, \phi_S} \exp\Bigl\{- \beta A[\eta _i] \Bigr\}\,\, \nonumber
\\
& &A[\eta_{i} ]=\, \sum_{i=1}^{N} \int_{0}^{\beta} d\tau \biggl[ \frac{\mu}{2} \dot{\eta }_i^2(\tau) + V_M[\eta_i(\tau)] + \,V_{sol}[\eta_i(\tau)] + V_{S}\bigl[ \eta _i(\tau), \eta _{i-1}(\tau) \bigr] \biggr] \, , \, 
\label{eq:7}
\end{eqnarray}

where $\mu=\,300 \, amu$ is the \textit{bp} reduced mass. The measure ${D}x_{i}$ is a multiple integral over the path Fourier coefficients while $\oint$ indicates that the paths $x_i(\tau)$ are closed trajectories. The integration over the two particles potential greatly enhances the computational times with respect to the previous method \cite{io14b} but it offers a more realistic model for the base stacking in sequences of any length.

\section{Free Energy of Minicircles}

There has been a widespread interest in the properties of small sequences of DNA following measurements of high cyclization probabilities which have pointed to an intrinsic flexibility for fragments of $\sim 100$ \textit{bps} or less \cite{cloutier}. It has been suggested that the formation of local disruptions in the helix of DNA loops may be the mechanism which permits to release the torsional stress and energetically favors the stability of bent molecules. Certainly, the breaking of some \textit{bps} and the opening of fluctuational bubbles change the average pitch of the helix, that is the number of \textit{bps} per helix turn.

\begin{figure}
\includegraphics[height=7.0cm,width=12.5cm,angle=-90]{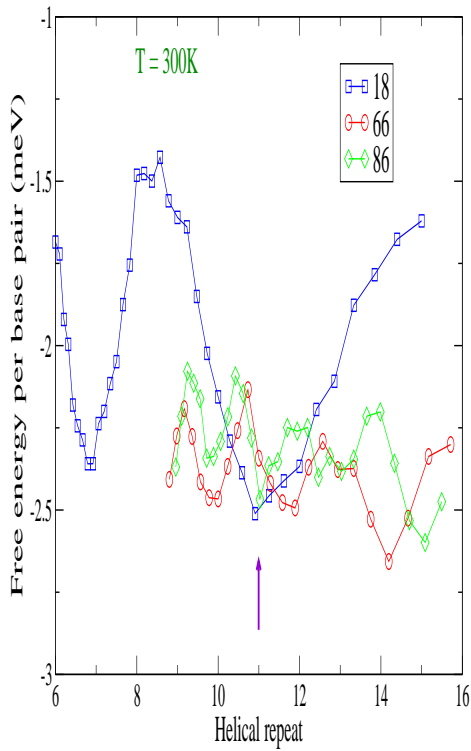}
\caption{\label{fig:2}(Color online) Free energy per base pair for three loops with $N=\,18, \,66, \,86$. The free energies are computed, at room temperature, as a function of the number of \textit{bps} per helix turn.}
\end{figure}

In the theoretical framework synthesized by Eq.~(\ref{eq:7}), I compute the free energy ($F=\,\beta^{-1}\ln Z_N$) of three heterogeneous loops with different length, $N=\,18, \,66, \,86$, but similar content of AT-\textit{bps} and GC-\textit{bps}, $\sim 50 \%$ each. The rise distance is pinned to the experimental value. The $66-$ and $86-$ \textit{bps} loops have been prepared as described in Ref.\cite{volo08} whereas the $N=\,18$ \textit{bps} is a toy sequence here introduced for comparison.  For each loop, we simulate a broad set of values for the twist $Tw$ and, for any $Tw$ (i.e., $h$), $Z_N$ is calculated by summing over an ensemble of fluctuational paths representing a large number of molecule configurations, about $10^7$ paths for each base pair in Eq.~(\ref{eq:7}).

Hence, the free energy is obtained as a function of the helical repeat $h$. 
The computational time for a simulation, e.g. for the $N=\,66$ sequence, is about eight days on a workstation (Intel Xeon E5-1620 v2, 3.7GHz processor).
The room temperature results for $F/N$ are plotted in Fig.~\ref{fig:2}. While the shortest loop shows a free energy minimum also at $h \sim 7$, the minima for all loops are remarkably found for $h$ in the range $\sim (10 - 12)$ in line with the well known values of helical pitch in DNA sequences \cite{calla}. Even more interestingly, the $66-$ and $86-$ minicircles also show free energy minima at larger $h$ due to  a spontaneous unwinding of the complementary strands. This $Tw$ reduction is consistent with the observations \cite{volo08} of \textit{bps} disruptions occuring in small loops as a consequence of the strong rotational deformations of the helix. The bending stress decreases by increasing the radius of the loop.

Then, for an ensemble of molecules with $N$ \textit{bps}, the free energy minimization evaluates the thermodynamically stablest values of helical repeat, that is an average property of the molecule ensemble. Furthermore, the path integral method can also determine the probabilities for the formation of fluctuational bubbles and select those base pairs along the DNA sequence for which hydrogen bond breaking is more likely to occur. By tuning the system temperature and the input parameters which control the counterions concentration in the solvent, we thus obtain a general and reliable computational scheme for the modeling of heterogeneous DNA loops in various ambient conditions.


\end{document}